\documentclass[aps,prfluids,11pt,tightenlines,longbibliography,superscriptaddress,onecolumn]{revtex4-2}
\usepackage{lineno}
\usepackage{amsmath,xcolor,graphicx,amssymb,bm,mathrsfs,amsthm,soul}
\newcommand{\rmd}{d}
\newcommand{\Req}{R_{\mathrm{eq}}}
\newcommand{\psieq}{\Psi_{\mathrm{eq}}}
\newcommand{\Pst}{P_{\mathrm{st}}}
\newcommand{\bfsOmega}{\bm{\mathsf{\Omega}}}
\newcommand{\OmegaOB}{\bm{\mathsf{\Omega}}^{\mathrm{OB}}}
\newcommand{\OmegaRPY}{\bm{\mathsf{\Omega}}^{\mathrm{RPY}}}
\newcommand{\OmegaZO}{\bm{\mathsf{\Omega}}^{\mathrm{ZO}}}
\newcommand{\Omegaeq}{\bm{\mathsf{\Omega}}_{\mathrm{eq}}}
\newcommand{\avOmega}{\langle\bfsOmega\rangle}
\newcommand{\sbfI}{\bm{\mathsf{I}}}
\newcommand{\Wi}{\mathrm{Wi}}
\newcommand{\Rmax}{R_\mathrm{max}}
\newcommand{\varpicav}{\varpi_{\mathrm{cav}}}

\begin{document}

\title{
% An analytical study of dumbbells with hydrodynamic interactions in fluctuating flows}
Turbulent stretching of dumbbells with hydrodynamic interactions: \\an analytical study}
% \title{An analytical study of dumbbells with hydrodynamic-interactions in fluctuating flows\\
% \title{Dumbbells with hydrodynamic-interactions in fluctuating flows: an analytical study}
% \title{How hydrodynamic interactions affect dumbbell stretching in fluctuating flows}

\author{Jason R. Picardo}
 \email{picardo@iitb.ac.in}
\affiliation{Department of Chemical Engineering, Indian Institute of Technology Bombay, Mumbai 400076, India}

\author{Dario Vincenzi} 
 \email{dario.vincenzi@univ-cotedazur.fr}
\affiliation{Universit\'e C\^ote d'Azur, CNRS, LJAD, 06100 Nice, France}

\begin{abstract}
  We study the stretching of an elastic dumbbell in a turbulent flow, with the aim of understanding and quantifying the effect of hydrodynamic interactions (HI) between the beads of the dumbbell. Adopting the Batchelor-Kraichnan model for the flow, we derive a Fokker-Planck equation and solve it analytically to obtain the probability distribution of the dumbbell's extension. Using different formulations of the HI tensor, we find that HI preferentially enhances the stretching of stiff dumbbells, i.e., those with a small Weissenberg number. We also evaluate the averaging approximations commonly used to simplify the description of HI effects; the consistently-averaged approximation shows that HI result in a less-pronounced coil-stretch transition in chaotic flows. Finally, we confirm the relevance of our analytical results by a comparison with Brownian dynamics simulations of dumbbells transported in a direct numerical simulation of homogeneous isotropic turbulence. 
\end{abstract}

\maketitle

\section{Introduction}

When a polymer is immersed in a fluid, the displacement of one of its parts disturbs the surrounding fluid and thus affects the motion of all other parts of the molecule.
Such hydrodynamic interactions (HI) between different parts of a polymer are especially important for long flexible molecules \cite{l05,s05,s18} and lie at the heart of phenomena such as conformation hysteresis in extensional flows~\cite{sbsc03} and migration effects in confined geometries~\cite{jendrejack04}. Moreover, the inclusion of HI in polymer models is now understood to be indispensable to achieve quantitative agreement with experiments~\cite{s18}.

Mesoscopic polymer models can easily incorporate effects such as finite extensibility, excluded volume forces, internal viscosity, and HI \cite{bird,l88,o96,g18}. In particular, in bead--spring or bead--rod chains,
HI are accounted for by assuming that the flow at the location of a given bead is the superposition of the background flow and the disturbances caused by the motion of all other beads that constitute the chain. Under the Stokes approximation, the velocity disturbance produced by a bead is a linear function of the hydrodynamic force that the bead exerts on the fluid.
The simplest bead--spring model is the elastic dumbbell. Without HI, the Hookean dumbbell model fails to capture important features of the non-Newtonian behavior of polymer solutions. But if HI are accounted for, the Hookean dumbbell in a shear flow exhibits shear-thinning behavior and yields shear-dependent normal-stress coefficients, with the second coefficient being negative \cite{fan86,fan87,diaz89,pp04}. 

Analytical solutions of the dumbbell model are available only for very simple flows, namely for steady potential linear flows \cite{bird}. The inclusion of HI makes the analytical study of the dumbbell model even more challenging. Several approximations have therefore been used to solve the dumbbell model with HI, namely closures of the evolution equation for the stress tensor \cite{ptt78}, various forms of averaging of the hydrodynamic interaction tensor \cite{zimm56,o85,w88,zo89}, and the Gaussian approximation of the configurational distribution function \cite{o89,zo89}. The latter approach has been shown to outperform the other approximation methods, since it accounts for the fluctuations of HI instead of averaging over them \cite{zo89,o96} (however, the addition of excluded-volume forces or internal viscosity limits the applicability of the Gaussian approximation \cite{pp02,kcp18}). The effects of HI have also been investigated using Brownian dynamics simulations and an extensive literature exists for steady flows---see Refs.~\cite{l05,s05,s18} for a review.

Single-polymer dynamics in turbulent flows has also been been well studied by means of Brownian dynamics simulations \citep{za03,gsk04,tdmsl04,jc07,wg10,bmpb12,vwrp21,ppv23,kg24}. However, these studies have typically ignored HI, with the exception of Refs.~\cite{sg03,kwt10,vwrp21}.
Here we aim to study the effect of fluctuating HI in turbulent flows analytically. 
In view of the complexity of the problem, we make some simplifying assumptions. We restrict ourselves to the dumbbell model with nonlinear elasticity. As a model for the turbulent strain-rate experienced by the dumbbell, we consider the Batchelor regime of the Kraichnan flow, where the velocity gradient is a statistically isotropic Gaussian tensor, with zero correlation time \cite{fgv01}.
The Kraichnan model is widely used in the analytical study of turbulent transport, with applications to turbulent mixing \cite{fgv01}, the small-scale fluctuating dynamo \cite{rincon19,tobias21}, and heavy particles in turbulent flows~\cite{bgm24}. In the context of single-polymer dynamics, the Batchelor-Kraichnan model has been used to calculate the exact form of the probability density function (PDF) of the extension of a polymer (without HI) in a turbulent flow: the PDF has a power-law tail with an exponent that increases from negative to positive values as the Weissenberg number (Wi) is increased (the exponent $-1$ marks the coil-stretch transition at Wi = 1/2) \cite{bfl00,c00,thiffeault03,mav05,pav16}. The model has also been used to estimate the scission time of a polymer \cite{vwrp21}. Although the assumptions of the Batchelor-Kraichnan model may seem rather restrictive, it has been found that the power-law behaviour of the PDF of the extension persists after the random flow is endowed with a correlation time comparable to that in homogeneous isotropic turbulence \cite{mv11}; the finite time-correlation does however introduce a local peak in the PDF near the maximum extension. A direct comparison of polymer stretching in a turbulent flow with that in a Gaussian flow (designed with a matching Lagrangian correlation time and Lyapunov exponent) reveals nearly the same statistics; the effect of the intermittent fluctuations of the turbulent velocity gradient remains rather mild at all Weissenberg numbers \cite{ppv23}. So, the the Batchelor-Kraichnan flow has proven to be a useful model, providing insight into the turbulent stretching of polymers; we show, towards the end of this paper, that this is true even with the addition of hydrodynamic interactions.
% the predictions of the Batchelor-Kraichnan flow do indeed anticipate the effect of HI on the turbulent stretching of dumbbells.

The dumbbell model is a minimal representation of a polymer molecule, and much more refined models are indeed available. It is therefore worth stressing the merits of a  study of this model.
It has already been mentioned that few analytical results are available when HI (without preaveraging) are incorporated in polymer models. While this applies to steady flows, it is even more relevant for fluctuating flows.
A theoretical result on the effect of HI in fluctuating flows therefore constitutes progress in our understanding of polymer dynamics. In addition, multiscale Eulerian--Lagrangian approaches are emerging as a promising tool for the numerical simulation of polymer solutions \cite{wg13,rll22,sbgc22}. In such approaches, a continuum description for the solvent is coupled with the Brownian dynamics of a suitable ensemble of polymer molecules.
Since a huge number of polymers, at least $10^8$, is required to obtain a realistic feedback on the flow, simulations have thus far used the most elementary version of the dumbbell model. Nonetheless, one of the advantages of the Eulerian--Lagrangian approach is that additional molecular-scale phenomena can be included in a straight-forward manner by refining the polymer model.
It is plausible, therefore, that HI will be incorporated in future Eulerian--Lagrangian simulations of turbulent polymer solutions, and so it is important to understand HI effects in detail. 
Finally, while HI play an important role in polymer dynamics, they generally produce small modifications in the statistics of polymer stretching. Their effect in turbulent flows may therefore be difficult to assess clearly in numerical simulations, owing to statistical errors. Thus, an analytically solvable model can yield useful insights into the effect of HI on single-polymer dynamics when the velocity gradient fluctuates randomly.

The rest of this paper is organized as follows. Section~\ref{sect:dumbbell} reviews the dumbbell model with HI and the most common forms for the hydrodynamic interaction tensor.
Section~\ref{sect:BK} introduces the random flow and derives the Fokker--Planck equation for the PDF of the end-to-end separation vector.
Section~\ref{sect:HI} examines the effect of HI on the statistics of polymer extensions and compares different forms of the hydrodynamic-interaction tensor. In Sect.~\ref{sect:averaging}, the consequences of averaging HI are analyzed. In Sect.~\ref{sect:dns}, the analytical predictions for the random flow are compared with Brownian dynamics simulations for a turbulent flow.
Finally, Sect.~\ref{sect:conclusions} concludes with a summary of the results, and a discussion of their implications for the numerical simulation of turbulent polymer solutions.

\section{The dumbbell model with hydrodynamic interactions}
\label{sect:dumbbell}

Consider an elastic dumbbell with bead radius $a$, spring coefficient $H$, and contour length $\Rmax$. The separation vector between the two beads is denoted as $\bm R(t)$.
The dumbbell is immersed in a Newtonian solvent of dynamic viscosity $\eta_s$ at temperature $T$; thus, the friction coefficient of each bead is $\zeta=6\pi\eta_s a$. 
We assume that the spring force $\bm F$ is given by the finitely extensible nonlinear elastic (FENE) force law, i.e.,
\begin{equation}
\label{eq:fene}
\bm F = \frac{H\bm R}{1-R^2/\Rmax^2}, \qquad R < \Rmax.
\end{equation}
The dumbbell's mean square length at equilibrium (in a still fluid)
is $R^2_{\mathrm{eq}}= 3(k_B T/H)(1+5/b)^{-1}$,
% = 3 \Rmax^2/(b+5)
where $b=H\Rmax^2/k_B T$ with $k_B$ the Boltzmann constant \cite{bird}. The dimensionless parameter $b$ measures the extensibility of the polymer. In the limit $b\to\infty$, the Hookean force law with spring coefficient~$H$ is recovered.

In turbulent flows, the polymer contour length is usually much smaller than the viscous scale, and the Reynolds number at the scale of the polymer is much smaller than unity.
The motion of the fluid surrounding the polymer is therefore given by a linear incompressible velocity field $\bm u(\bm x,t)$ with gradient $\bm\kappa(t)$ such that $\kappa_{ij}=\partial u_i/\partial x_j$, $i,j=1,2,3$. Likewise, hydrodynamic interactions between the beads are accounted for by assuming that the velocity disturbance near one bead due to the motion of the other bead is given by the Stokes flow.

%For a deterministic $\bm\kappa(t)$, 
Under the above assumptions, the probability density function (PDF) of $\bm R$, denoted by $\Psi(\bm R;t)$, satisfies the advection--diffusion equation~\cite{l88,o96}
\begin{equation}
\label{eq:CD}
\partial_t \Psi =
-\nabla_{\bm R}\cdot \big[\bm{\kappa}(t)\cdot \bm R \,\Psi
  -\frac{2}{\zeta} (\bm{\mathsf{I}}-\zeta\bm{\mathsf{\Omega}})\cdot\bm F\,\Psi\big]
  +\frac{2k_BT}{\zeta}\nabla_{\bm R}\cdot(\bm{\mathsf{I}}-\zeta\bfsOmega)\cdot\nabla_{\bm R}\Psi,
\end{equation}
where $\bm{\mathsf{I}}$ is the identity tensor and $\bfsOmega$  the hydrodynamic-interaction tensor. We shall consider different forms of $\bfsOmega$, the simplest being the Oseen--Burgers (OB) tensor
\begin{equation}
  \OmegaOB = \frac{1}{8\pi\eta_s R}\left(\bm{\mathsf{I}} + \frac{\bm R \bm R}{R^2}\right)
  = \frac{3}{4\zeta R}\,h^*\,
  \sqrt{\frac{\pi k_B T}{H}}\left(\bm{\mathsf{I}} + \frac{\bm R \bm R}{R^2}\right),
\end{equation}
where $h^*=a/\sqrt{\pi k_BT/H}$ is the hydrodynamic-interaction parameter. Its maximum physically-consistent value corresponds to setting $a = R_\mathrm{eq}/2$ which yields $h^* \approx 0.49$; it is commonly taken in the range $0\leqslant h^*\lesssim 0.3$. A limitation of the OB formulation is that the diffusion tensor in Eq.~\eqref{eq:CD}, $(2k_BT/\zeta) (\bm{\mathsf{I}}-\zeta\bfsOmega)$, is not positive-semidefinite for $R<3a/2$ \cite{o96}.
The lack of positiveness of $\OmegaOB$ for small separations can be overcome by considering the Rotne--Prager--Yamakawa (RPY) tensor \cite{rp69,y70}
\begin{equation}
  \OmegaRPY =
 \frac{3}{4\zeta R}\,h^*\,
  \sqrt{\frac{\pi k_B T}{H}}
  \begin{cases}
    \dfrac{R}{2a}\left[\left(\dfrac{8}{3}-\dfrac{3R}{4a}\right)\bm{\mathsf{I}} + \dfrac{R}{4a}\dfrac{\bm R\bm R}{R^2}\right] & (R <  2a), 
   \\[15pt]
    \left(1+\dfrac{2a^2}{3R^2}\right)\bm{\mathsf{I}} + \left(1-\dfrac{2a^2}{R^2}\right)\dfrac{\bm R \bm R}{R^2} &  (R\geqslant 2a).
   \end{cases}
\end{equation}
$\OmegaRPY$ is currently the most common choice for the hydrodynamic-interaction tensor in bead--spring models of polymers.
To avoid the inconvenience of dealing with a piecewise-defined tensor in analytical calculations and higher-order integration schemes, Zylka and \"Ottinger \cite{zo89} proposed the following approximation of the  RPY tensor:
\begin{equation}
\OmegaZO = \frac{3h^*}{4\zeta R(R^2+4a^2/3)^3}
\sqrt{\frac{\pi k_B T}{H}}
\left[
\left( R^6 + \frac{14}{3}\,a^2 R^4 + 8\,a^4 R^2\right) \sbfI +
\left( R^6 + 2\,a^2R^4 - \frac{8}{3}\, a^4R^2\right) \dfrac{\bm R \bm R}{R^2}
\right].
\end{equation}
The tensor $\OmegaZO$ is positive-definite, regular at $R=0$, and coincides with $\OmegaOB$ to order $R^{-1}$ and with $\OmegaRPY$ to order $R^{-3}$ as $R\to\infty$; furthermore, it captures the effect of finite bead size to order $R^{-5}$ \cite{zo89}.

Equation~\eqref{eq:CD} can be recast as a Fokker--Planck equation:
\begin{equation}
\label{eq:FPE-laminar}
\partial_t \Psi = -\frac{\partial}{\partial R_i}\big[\kappa_{ij}(t)R_j
  -\frac{2}{\zeta} (\delta_{ij}-\zeta\mathsf{\Omega}_{ij})F_j-2k_BT\frac{\partial\mathsf{\Omega}_{ij}}{\partial R_j}\big]\Psi
  +\frac{2k_BT}{\zeta}\frac{\partial}{\partial R_i}\frac{\partial}{\partial R_j}(\delta_{ij}-\zeta\mathsf{\Omega}_{ij})\Psi,
\end{equation}
where summation over repeated indices is assumed.
The associated It\^o stochastic differential equation for $\bm R(t)$ is
\begin{equation}
  d\bm R = \bm\kappa(t)\cdot\bm R\,d t
  -\frac{2}{\zeta}(\bm{\mathsf{I}}-\zeta\bm{\mathsf{\Omega}})\cdot\bm F\,dt 
  - 2k_BT(\nabla_{\bm R}\cdot\bm{\mathsf{\Omega}})\, dt +\sqrt{\frac{4k_BT}{\zeta}}\,\bm{\mathsf{B}} \cdot d\bm W(t),
  \label{eq:SDE}
\end{equation}
where $\bm W(t)$ is the three-dimensional Brownian noise and $\bm{\mathsf{B}}$ is any matrix such that $\bm{\mathsf{B}}\bm{\mathsf{B}}^\top=(\bm{\mathsf{I}}-\zeta\bm{\mathsf{\Omega}})$
\cite{o96}. The form of $\bm{\mathsf{B}}$ will in fact be irrelevant in what follows, since we will focus on the probability density function of $R$ and hence on the Fokker--Planck equation. Also note that, in both Eqs.~\eqref{eq:FPE-laminar} and ~\eqref{eq:SDE}, the drift term proportional to $\nabla_{\bm R}\cdot\bm{\mathsf{\Omega}}$ vanishes for all the hydrodynamic-interaction tensors considered here,
owing to the incompressibility of the solvent \cite{o96}. 

\section{The Batchelor regime of the Kraichnan flow}
\label{sect:BK}

In the Batchelor regime of the Kraichnan model, $\bm\kappa(t)$ is a statistically isotropic Gaussian tensor with zero mean and zero correlation time \cite{fgv01}. The statistics of $\bm\kappa(t)$ is therefore entirely specified by its two-time correlation, which takes the form
\begin{equation}
\langle \kappa_{mn}(t) \kappa_{pq}(t') \rangle = \mathsf{K}_{mnpq} \delta(t-t')
\end{equation}
with $m,n,p,q=1,2,3$ and
\begin{equation}\label{eq:BKgrad}
 \mathsf{K}_{mnpq} = \frac{\lambda}{3}\, (4 \delta_{mp}\delta_{nq} - \delta_{mn}\delta_{pq}-\delta_{mq}\delta_{np}).
\end{equation}
Here $\lambda$ is the maximum Lyapunov exponent of the flow, i.e., the average exponential rate of separation of nearby fluid particles. It therefore provides a measure of stretching in the fluctuating flow.

The randomly fluctuating $\bm\kappa(t)$ in Eq.~\eqref{eq:SDE} plays the role of a multiplicative tensorial white noise. The Stratonovich and It\^o interpretations of the stretching term $\bm\kappa(t)\cdot\bm R\,dt$ is identical in the present case, since the flow is incompressible (see the appendix in Ref.~\cite{fgv01}). So, regarding Eq.~\eqref{eq:SDE} as an It\^o stochastic differential equation, and
% There are now two sources of randomness in Eq.~\eqref{eq:SDE}. Let us denote the associated probability density function of $\bm R$ at time $t$ as $f(\bm R;t)$.
extending the rules of stochastic calculcus to a tensorial noise (see, e.g., Ref.~\cite{pav16}), we obtain the following Fokker--Planck equation for the PDF of $\bm R$, denoted by $f(\bm R;t)$:
\begin{equation}
  \label{eq:FPE-random}
  \partial_t f = -\dfrac{\partial}{\partial R_i}(V_i f) + \frac{1}{2}\frac{\partial}{\partial R_i}\frac{\partial}{\partial R_j}(D_{ij}f), \quad i,j=1,2,3,
\end{equation}
with drift and diffusion coefficients
\begin{flalign}
  \label{eq:Vi}
  V_i &= -\frac{2}{\zeta} (\delta_{ij}-\zeta\mathsf{\Omega}_{ij})F_j-2k_B T\frac{\partial\mathsf{\Omega}_{ij}}{\partial R_j} + \frac{1}{2}\mathsf{K}_{ijjk}R_k =
  -\frac{2}{\zeta} (\delta_{ij}-\zeta\mathsf{\Omega}_{ij})F_j,
  %A_i + \frac{1}{2}\mathcal{K}_{ijjk}q_k,
  \\
  D_{ij} &= \frac{4k_B T}{\zeta}(\delta_{ij}-\zeta\mathsf{\Omega}_{ij}) + \mathsf{K}_{ikjl}R_k R_l.
\end{flalign}
In Eq.~\eqref{eq:Vi}, we have used the fact that $\bfsOmega$ is divergenceless and $\mathsf{K}_{ijjk}=0$, a consequence of the flow incompressibility.

Taking advantage of the statistical isotropy of the flow, we move to spherical coordinates $\bm R = (R\sin\theta\cos\varphi, R\sin\theta\sin\varphi,R\cos\theta)$ with $0\leqslant\theta\leqslant\pi$, $0\leqslant\varphi\leqslant2\pi$, and $0\leqslant R< \Rmax$ for all choices of $\bfsOmega$ except for the OB tensor, for which $3a/2 \leqslant R <\Rmax$.
The probability density function $P(R,\theta,\varphi;t)$ satisfies a new Fokker--Planck equation whose drift and diffusion coefficients are calculated from $V_i$ and $D_{ij}$ according to the transformation rules of a Fokker--Planck equation under a change of variables (see Ref.~\cite{r89}, p.~84).
In the case under consideration, the Fokker--Planck equation for $P(R,\theta,\varphi;t)$ takes the form
\begin{equation}
  \label{eq:FPE_spherical}
\partial_t P = -\partial_R (V_R P) - \partial_\theta (V_\theta P) + \frac{1}{2}\partial_R^2 (D_{RR}P) + \frac{1}{2}\partial_\theta^2 (D_{\theta\theta}P) + \frac{1}{2}\partial_\varphi^2 (D_{\varphi\varphi}P),
\end{equation}
where 
\begin{gather}
  V_R = \frac{4\lambda R}{3} - \frac{2HR}{\zeta(1-R^2/L^2)} \, \mathfrak{G}_1 + \frac{4k_B T}{\zeta R}\, \mathfrak{G}_2,
  \quad
  V_\theta = \left(\frac{2\lambda}{3} + \frac{2k_B T}{\zeta R^2} \, \mathfrak{G}_2\right)\cot\theta,
  \\[10pt]
  D_{RR}= \frac{2\lambda R^2}{3} + \frac{4k_B T}{\zeta}\, \mathfrak{G}_1,
  \quad
  D_{\theta\theta} = \frac{4\lambda}{3} + \frac{4k_B T}{\zeta R^2}\, \mathfrak{G}_2,
  \quad
  D_{\varphi\varphi} = \left(\frac{\lambda}{3} + \frac{k_B T}{\zeta R^2}\, \mathfrak{G}_2\right)\frac{4}{\sin^2\theta},
\end{gather}
while $V_\varphi$, $D_{R\theta}$, $D_{R\varphi}$, $D_{\theta\varphi}$ are all zero, in keeping with the statistical isotropy of the flow. The coefficients $\mathfrak{G}_1$ and $\mathfrak{G}_2$ depend on the choice of the hydrodynamic-interaction tensor. We have, for the OB tensor,
\begin{equation}
\mathfrak{G}_1 = 1-\frac{3a}{2R}, \qquad \mathfrak{G}_2 = 1-\frac{3a}{4R},
\end{equation}
while for the RPY tensor,
\begin{equation}
  \mathfrak{G}_1=
  \begin{cases}
    \dfrac{3R}{16a} & (R < 2a),
    \\[10pt]
    1-\dfrac{3a}{2R}+\left(\dfrac{a}{R}\right)^3 & (R\geqslant 2a),
  \end{cases}
\qquad
  \mathfrak{G}_2=
  \begin{cases}
    \dfrac{9R}{32a} & (R < 2a),
    \\[10pt]
    1-\dfrac{3a}{4R}-\dfrac{1}{2}\left(\dfrac{a}{R}\right)^3 & (R\geqslant 2a),
  \end{cases}
\end{equation}
and for the ZO tensor, 
\begin{align}
  \mathfrak{G}_1 &= 1-\dfrac{3a}{2R}+\left(\dfrac{a}{R}\right)^3 - \frac{16a^7}{R^3(4a^2+3R^2)^2},
  \\[10pt]
  \mathfrak{G}_2 &= 1-\dfrac{3a}{4R}-\dfrac{1}{2}\left(\dfrac{a}{R}\right)^3 + \frac{8(4a^9+15a^7R^2)}{(4a^2R+3R^3)^3}.
\end{align}
Since the velocity gradient is statistically isotropic, the stationary PDF of $\bm R$ is sought in the form $P(R, \theta, \varphi;t) = \Pst(R)\sin\theta$.
For all choices of $\bfsOmega$, replacing this form in Eq.~\eqref{eq:FPE_spherical} yields the following equation for $\Pst(R)$:
\begin{equation}
  \label{eq:Pst}
  \partial_R(V_R \Pst) = \frac{1}{2}\partial^2_R(D_{RR}\Pst).
\end{equation}
We impose reflecting boundary conditions at the endpoints of the range of $R$. This means that the probability current $J=2V_R\Pst-\partial_R(D_{RR}\Pst)$ vanishes for all $R$, and the solution of Eq.~\eqref{eq:Pst} takes the form \cite{r89}
\begin{equation}
\label{eq:pst_integral}
\Pst(R) = \frac{N}{D_{RR}} \exp\left(2\int_{R_\star}^R \frac{V_R(\rho)}{D_{RR}(\rho)}\,\rmd\rho\right),
\end{equation}
where the value of $R_\star$ is irrelevant, as it only modifies the normalization coefficient $N$, which must satisfy the normalization condition
\begin{equation}  
\label{eq:normal}\int_0^{\Rmax}\int_0^\pi\int_0^{2\pi} \Pst(R)\sin\theta\,\rmd R\,\rmd\theta\,\rmd\varphi = 4\pi \int_0^{\Rmax}\Pst(R)\rmd R =1.
\end{equation}
Note that, in the case of the OB tensor, the lower bound in the range of $R$ in Eq.~\eqref{eq:normal} should obviously be $3a/2$ instead of $0$. 

We recall that, for a free-draining dumbbell ($h^*=0$), the integral in Eq.~\eqref{eq:pst_integral} can be calculated exactly to obtain
(see Refs.~\cite{c00,thiffeault03,mav05})
\begin{equation}
\label{eq:pdf_hstar_0}
\Pst(R) = N R^2 \left( 1 + \frac{2b\Wi R^2}{3\Rmax^2} \right)^{-\gamma} \left(1-\frac{R^2}{\Rmax^2}\right)^\gamma,
\end{equation}
where $\Wi=\lambda\tau$ is the Weissenberg number,  with $\tau=\zeta/4H$ the elastic relaxation time. Also, $\gamma^{-1} = {2}/{3}\left(3b^{-1}+2\Wi\right)$,
% \begin{equation}
% \gamma^{-1} = \frac{2}{3}\left(3b^{-1}+2\Wi\right),
% \end{equation}
and
\begin{equation}
\label{eq:N}
N^{-1} = \frac{\pi^{3/2}\, \Rmax^3\Gamma(\gamma+1)}{\Gamma(5/2+\gamma)} \, _2 F_1(3/2,\gamma;5/2+\gamma;-2b\Wi/3),
\end{equation}
with $_2 F_1$ denoting the hypergeometric function. In particular, note that $\Pst(R)\sim R^{2(1-\gamma)}$ for $\Req \ll R\ll \Rmax$. The power-law behavior of the PDF for intermediate extensions is a general result for fluctuating flows~\cite{bfl00,ppv23}.

The knowledge of the PDF of $R$ allows the calculation of the polymeric contribution to the mean stress tensor, which, according to the Kramers expression \cite{bird}, is  
\begin{equation}
\boldsymbol{\mathsf{T}}_p = -n\langle\bm R\,\bm F\rangle,
\end{equation}
where $n$ is the number density of the polymers and the brackets denote the average over the stationary PDF of $\bm R$. As a consequence of statistical isotropy, 
\begin{equation}
\bm{\mathsf{T}}_p = \frac{4\pi}{3} nH \overline{\mathsf{T}}\, \bm{\mathsf{I}}
\end{equation}
with
\begin{equation}
\label{eq:Tbar}
\overline{\mathsf{T}}=\int_0^{\Rmax} \frac{R^2}{1-R^2/\Rmax^2}\, \Pst(R)\rmd R.
\end{equation}

\section{Effect of hydrodynamic interactions on extension}
\label{sect:HI}

\begin{figure}
  \centering
 \includegraphics[width=0.495\textwidth]{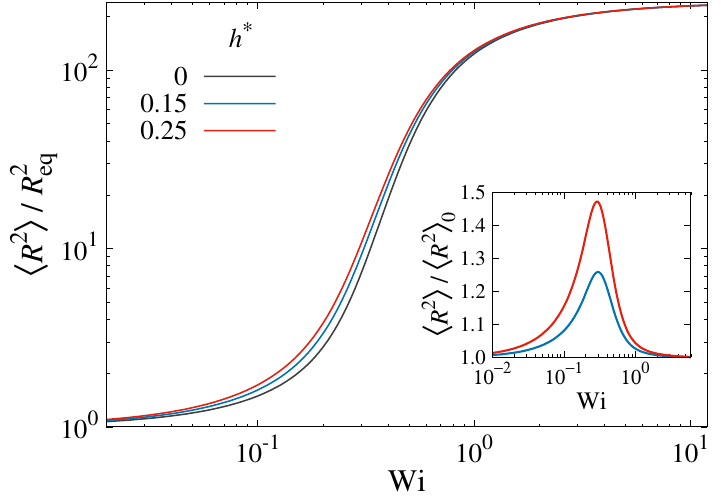}%
 \put(-193,39){\footnotesize (a)}
  \put(38,39){\footnotesize (b)}
   \hfill%
 \includegraphics[width=0.505\textwidth]{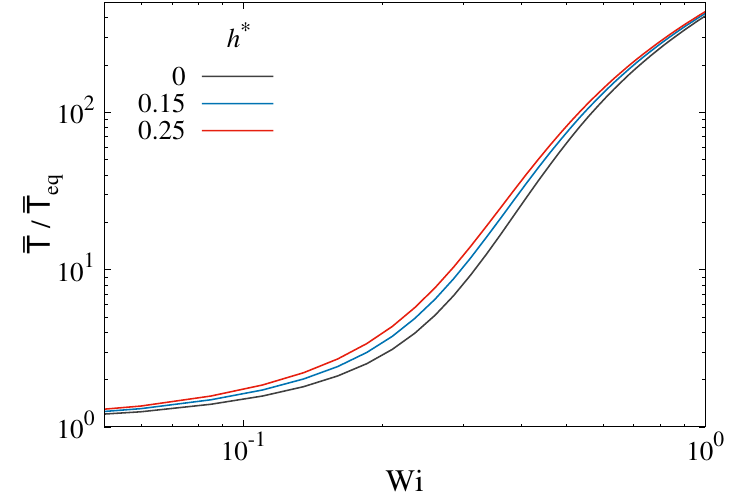}%
  \caption{(a) Increase of the polymer mean square extension as a function of the Weissenberg number, for different values of the hydrodynamic-interaction parameter. The inset shows the mean square extension rescaled with its value for $h^*=0$, denoted as $\langle R^2\rangle_0$; the color code is the same as in the main plot. (b) Increase of the mean polymer stress as a function of the Weissenberg number, for different values of the hydrodynamic-interaction parameter. The RPY hydrodynamic-interaction tensor is used for both panels and the extensibility parameter is $b=1200$. The scaling factors $R^2_\mathrm{eq}$ and $\overline{\mathsf{T}}_\mathrm{eq}$ are the equilibrium values of the mean square extension and the polymeric stress.}
  \label{fig:RPY_hstar}
\end{figure}

We first
assess the effect of HI on the statistics of the extension and the mean polymer stress. 
To this end, we calculate the stationary PDF of the extension from Eqs.~\eqref{eq:pst_integral} and \eqref{eq:normal}.
We then obtain the rescaled mean square extension $\langle R^2\rangle/\Req^2$ and 
the magnitude of the mean polymer stress [from Eq.~\eqref{eq:Tbar}].
The extensibility parameter is set to $b=1200$, and HI are described by the RPY hydrodynamic tensor with
$h^*=0,0.15,0.25$, where $h^*=0$ corresponds to a free-draining dumbbell (no HI). This choice of parameters is consistent with previous studies of dumbbells in laminar flows \cite{o85,fan86,fan87,w88,diaz89,zo89,pp02,kcp18}.

Figure~\ref{fig:RPY_hstar}(a) shows the relative increase in the polymer mean square extension as a function of the Weissenberg number. The inclusion of HI in the dumbbell model results in a slightly increased polymer stretching, which is most pronounced near the coil--stretch transition [inset of Fig.~\ref{fig:RPY_hstar}(a)]. 
This effect of HI, in a fluctuating flow, is consistent with that in steady shear flows, where a similar magnitude of increased stretching is observed (see Refs.~\cite{fan86,fan87,diaz89,w88}). Along with the extension, the polymeric contribution to the mean stress also increases when $h^*>0$ [Fig.~\ref{fig:RPY_hstar}(b)].
% Such increase is consistent, also in magnitude, with that observed for steady shear flows (see Refs.~\cite{fan86,fan87,diaz89,w88}). A corresponding enhancement of the polymeric contribution to the stress is observed when $h^*>0$ [Fig.~\ref{fig:RPY_hstar}(b)].

\begin{figure}
  \centering
\includegraphics[width=0.505\textwidth]{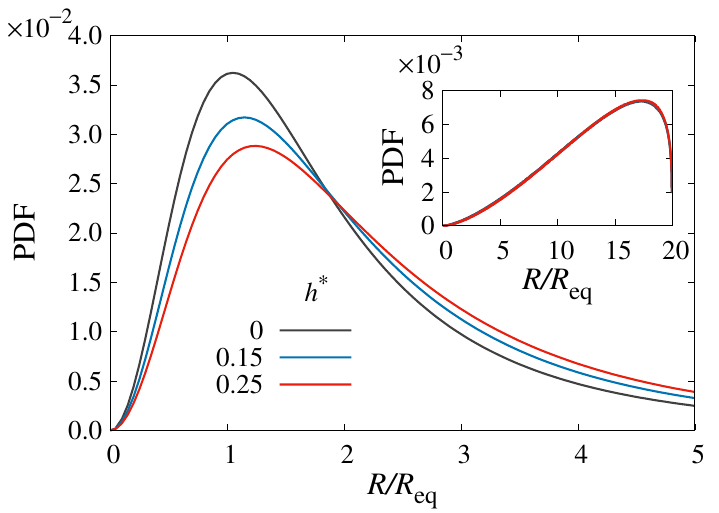}%
\put(-195,145){\footnotesize (a)}
% \put(41,132){\footnotesize (b)}
\put(41,35){\footnotesize (b)}
\hfill%
\includegraphics[width=0.495\textwidth]{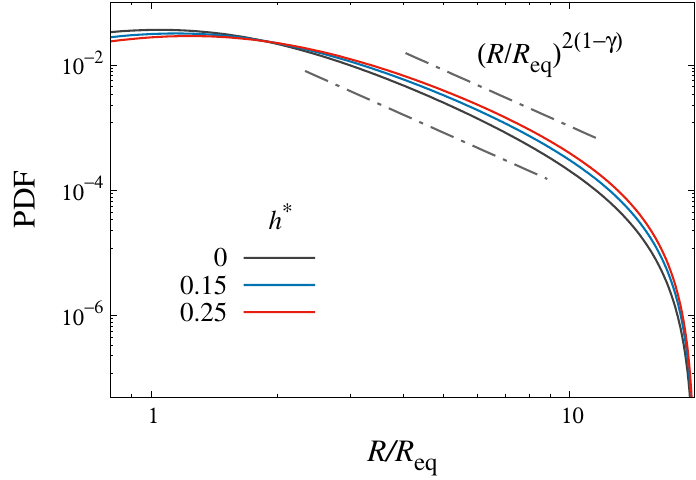}%
  \caption{(a) Stationary PDF of the polymer extension for different values of $h^*$ and $\Wi=0.3$ (main plot) and $\Wi=3.0$ (inset). The RPY hydrodynamic-interaction tensor is used to model HI and the extensibility parameter is $b=1200$. To facilitate a comparison, the PDFs are plotted over a limited range of $R$.
  (b) The same PDFs as in the main plot of panel (a) in double logarithmic scale. The dashed lines indicate the slope of the PDF for $\Wi=0.3$ and  $h^*=0$ as given by Eq.~\eqref{eq:pdf_hstar_0}.  }
  \label{fig:pdf_RPY}
\end{figure}

The effects of HI on the mean extension are a consequence of modifications to the stationary PDF of the extension. These changes are illustrated in Fig.~\ref{fig:pdf_RPY} (see the main panel which plots the PDF for $\Wi=0.3$). HI deplete the probability of near-equilibrium extensions, while they increase the probability of intermediate extensions. Hi effects, of course, vanish at large extensions for which the separation between the beads is much larger than their hydrodynamic radii. Consequently, HI does not alter the PDF of the extension when Wi is large, since the majority of polymers are strongly stretched---see the inset of Fig.~\ref{fig:pdf_RPY}(a), wherein the PDF for Wi = 3.0 is unaffected by the value of $h^*$. Despite these alterations in the PDF, its power-law behavior at intermediate extensions is preserved, as seen in Fig.~\ref{fig:pdf_RPY}(b) which plots the PDF on double logarithmic axes; indeed, the power-law exponent is largely unaffected by HI.

\begin{figure}
  \centering
  \includegraphics[width=0.5\textwidth]{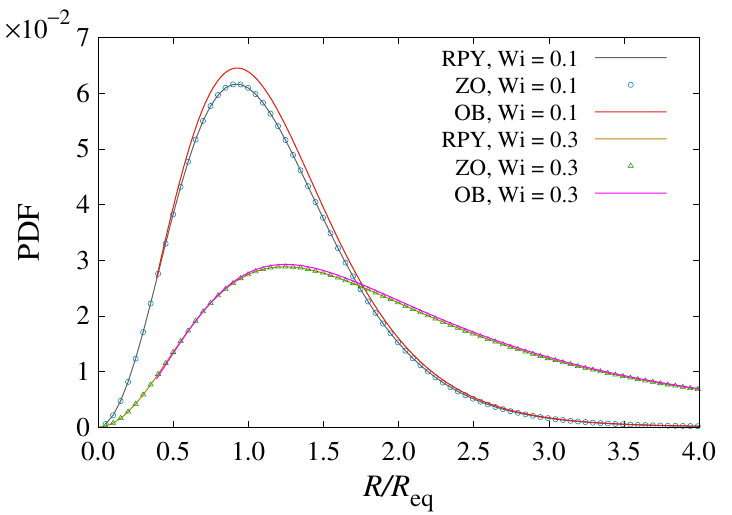}%
  \put(-197,140){\footnotesize (a)}
  \put(39,140){\footnotesize (b)}
  \hfill%
  \includegraphics[width=0.5\textwidth]{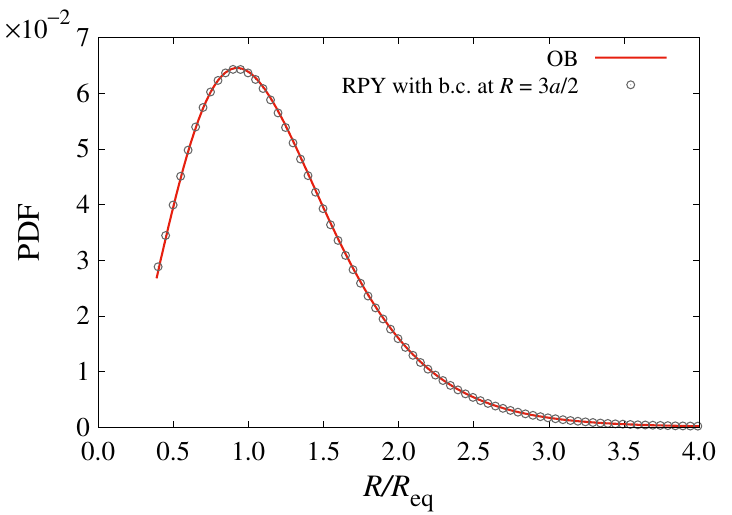}%
  \caption{(a) Stationary PDF of the extension for $b=1200$, $h^*=0.25$, $\Wi=0.1,0.3$, and
    different forms of the hydrodynamic-interaction tensor. For the OB tensor,
    $3a/2 \leq R \leq \Rmax$. 
    (b) Stationary PDF of the extension for the OB and RPY tensors, when reflecting boundary conditions are applied at  $R=3a/2$ for both cases. The parameters for this panel are $b=1200$, $h^*=0.25$, $\Wi=0.1$.}
  \label{fig:Omega}
\end{figure}

In Sect.~\ref{sect:dumbbell}, we recalled three forms of the hydrodynamic-interaction tensor often used in the literature to model HI in bead--spring models. Each of these forms results in different expressions for the drift and diffusion coefficients of the Fokker--Planck equation \eqref{eq:FPE_spherical}. To examine how the choice of the hydrodynamic-interaction tensor affects dumbbell stretching, we plot the PDF of $R/\Req$ for the different forms of $\bfsOmega$, in Fig.~\ref{fig:Omega}. We find that the PDFs for the RPY and ZO tensors coincide. The PDF for the OB tensor deviates slightly from the other two at small extensions and only for very small Wi. To understand this deviation, recall that using the OB tensor requires the reflecting boundary condition, for the Fokker--Planck equation, to be applied at $R=3a/2$ (since the OB tensor looses positive definiteness for smaller $R$). When we calculate the PDF for the RPY tensor by applying the reflecting boundary condition at $R=3a/2$ (instead of $R=0$), we obtain a result that is indistinguishable from the PDF for the OB tensor [see Fig.~\ref{fig:Omega}(b)].
Therefore, the deviation observed in Fig.~\ref{fig:Omega}(a) is only due to 
a difference in the normalization constant $N$, which is in turn caused by the difference in the range of the variable $R$.
This means, in particular, that the three forms of the hydrodynamic-interaction tensor would yield similar results if the two beads of the dumbbell were not allowed to overlap, i.e., $R>2a$ (see, for instance, the Brownian dynamics simulations in Ref.~\cite{diaz89}).
In any case, the effect of the different boundary condition becomes negligible for $\Wi \gtrsim 0.3$, since the contribution of the small extensions to the normalization constant of the PDF decreases with Wi.

The above analysis shows that the specific approximation for the hydrodynamic-interaction tensor has little impact on the statistics of polymer stretching, a conclusion consistent with the results obtained previously for steady flows \cite{diaz89,zo89}.

\section{Averaged hydrodynamic interactions}
\label{sect:averaging}

Our understanding of the effects of HI in steady viscometric flows owes much to the use of averaged approximations for the HI tensor $\bfsOmega$, which facilitated the analytical study of HI in both dumbbell and chain polymer models \citep{o96}. In this section, we leverage our exact knowledge of the PDF of the extension in the Bathelor-Kraichnan flow to examine how the validity of averaging approximations is impacted by the fluctuating flow.

The simplest of these approximations is Zimm's ``preaveraging approximation"
 \cite{zimm56}, which consists of replacing the fluctuating tensor
 $\bfsOmega$ in Eq.~\eqref{eq:CD}  with its equilibrium average
\begin{equation}
  \Omegaeq = \int \bfsOmega\, \psieq(\bm R) \rmd\bm R,
\end{equation}
where $\psieq(\bm R)$ is the solution of Eq.~\eqref{eq:CD} for $\bm\kappa=0$ \cite{note-equilibrium}. The preaveraged tensor
$\Omegaeq$ is necessarily proportional to the identity tensor, since $\psieq(\bm R)$ can only depend on $R$. 
For the FENE model, 
\begin{equation}
\psieq(\bm R)= \frac{1}{2\pi\Rmax^3 \mathrm{B}(3/2,1+b/2)}  \left(1-\frac{R^2}{\Rmax^2}\right)^{b/2},
\end{equation}
where B is the beta function. 
If, for simplicity, we use the OB approximation for the hydrodynamic-interaction tensor, we find
\begin{equation}
  \Omegaeq = \varpi_{\rm eq} \bm{\mathsf{I}}, \qquad
  \varpi_{\rm eq} = \sqrt{\frac{\pi}{b}}\, \frac{h^*}{\zeta (b/2+1)\mathrm{B}(3/2,b/2+1)}.
\end{equation}
Clearly, $\Omegaeq$ does not depend on $R$ and is positive definite. Therefore, when $\bfsOmega$ is replaced with
$\Omegaeq$ the PDF of $R$ is defined down to $R=0$, even in the OB approximation.

It is easily seen from Eqs.~\eqref{eq:CD} and \eqref{eq:SDE} that preaveraged hydrodynamic interactions simply modify the friction coefficient. Thus,  Eqs.~\eqref{eq:CD} and \eqref{eq:SDE} become equivalent to the equations for a free-draining dumbbell with friction coefficient $\tilde{\zeta}=\zeta/(1-\zeta\varpi_{\rm eq})$ \cite{bird}.
In other words, the relaxation time $\tau=\zeta/4H$ is modified to $\tilde{\tau}=\tilde{\zeta}/4H=\tau/(1-\zeta\varpi_{\rm eq})$, and preaveraged hydrodynamic interactions result in an increase of the effective Weissenberg number by a factor of $(1-\zeta\varpi_{\rm eq})^{-1}$, for \textit{all} Wi. The limit of $\zeta\varpi_{\rm eq}$ for $b\gg 1$ is $\sqrt{2}h^*$. For typical values of $b$ and $h^*$, therefore, the effective Weissenberg number is roughly
$1.5$ times greater than the actual Wi, regardless of the magnitude of Wi. Consequently, the preaveraging approximation yields the spurious prediction that the effect of HI on dumbbell stretching persists even at high Wi.

To improve upon the preaveraging approximation, \"Ottinger \cite{o85} proposed the ``consistently averaged approximation''. Here, instead of $\psieq(\bm R)$, one uses the average of $\bfsOmega$ over the stationary solution of Eq.~\eqref{eq:CD}, $\Psi_{\mathrm{st}}(\bm R)$. Since $\Psi_{\mathrm{st}}(\bm R)$ is unknown a priori, one begins with an ansatz for $\avOmega$ that contains free parameters. $\Psi_{\mathrm{st}}(\bm R)$ is then obtained by replacing $\bfsOmega$ with $\avOmega$ in Eq.~\eqref{eq:CD}. Finally, the average of $\bfsOmega$ is recalculated, using $\Psi_{\mathrm{st}}(\bm R)$, and then compared with the original guess for $\avOmega$ to determine the free parameters.

In the Batchelor--Kraichnan model, the velocity gradient is statistically isotropic. Therefore, the stationary PDF of $\bm R$ is necessarily of the form 
$P(R,\theta,\varphi)=\Pst(R)\sin\theta$, and
the average tensor  $\avOmega$ must once again be proportional to the identity tensor:
\begin{equation}
\langle\bfsOmega\rangle = \varpicav\, \bm{\mathsf{I}},  
\end{equation}
where $\varpicav$ remains to be determined. 
On replacing $\bfsOmega$ with $\varpicav\, \bm{\mathsf{I}}$ in Eq.~\eqref{eq:FPE-random},
we find that $\Pst(R)$ has the same form as in Eqs.~\eqref{eq:pdf_hstar_0} and \eqref{eq:N}, but with Wi replaced by 
\begin{equation}\label{eqn:effWi}
\widetilde{\mathrm{Wi}} = \frac{\mathrm{Wi}}{1-\zeta\varpicav}.
\end{equation}
%
%is $P(R,\theta,\varphi)=\Pst(R)\sin\theta$ with
%\begin{equation}
%\Pst(R) = N R^2 \left( 1 + \frac{2\widetilde{\mathrm{Wi}} R^2}{\Req^2} \right)^{-\gamma} \left(1-\frac{R^2}{L^2}\right)^\gamma,
%\end{equation}
%where
%\begin{equation}
%\widetilde{\mathrm{Wi}} = \frac{\mathrm{Wi}}{1-\varpi\zeta},
%\qquad
%\gamma^{-1} = \frac{2}{3}\left(\frac{\Req^2}{L^2}+2\widetilde{\mathrm{Wi}}\right),
%\end{equation}
%and
%\begin{equation}
%N(\Req,L,\widetilde{\Wi})^{-1} = \frac{\pi^{3/2}\, L^3\Gamma(\gamma+1)}{\Gamma(5/2+\gamma)} \, _2 F_1(3/2,\gamma;5/2+\gamma;-2b\widetilde{\mathrm{Wi}}/3)
%\end{equation}
%with $_2 F_1$ denoting the hypergeometric function. 
 The Wi-dependent value of $\varpicav$ is calculated by requiring
\begin{equation}
\varpicav \bm{\mathsf{I}} = \int_0^{\Rmax}\int_0^\pi\int_0^{2\pi} \bfsOmega\, \Pst(R)\sin\theta\,\rmd R\,\rmd\theta\,\rmd\varphi.
\end{equation}
For the OB hydrodynamic-interaction tensor, we have
\begin{equation}
\varpicav \bm{\mathsf{I}} = \frac{3}{4\zeta R}\,h^*\,
  \sqrt{\frac{\pi k_B T}{H}} \int_0^{\Rmax}\int_0^\pi\int_0^{2\pi} \left(\bm{\mathsf{I}} + \frac{\bm R \bm R}{R^2}\right) \Pst(R)\sin\theta\,\rmd R\,\rmd\theta\,\rmd\varphi,
\end{equation}
and the following implicit equation for $\varpicav$ obtains:
\begin{equation}
\varpicav = \frac{2h^*}{\sqrt{b}\,\zeta}\, \frac{\Gamma(5/2+\gamma)}{\Gamma(2+\gamma)}\,
\frac{_2F_1(1,\gamma;2+\gamma;-2b\widetilde{\rm Wi}/3)}{_2F_1(3/2,\gamma;5/2+\gamma;-2b\widetilde{\Wi}/3)}.
\end{equation}
\begin{figure}
  \centering
  \includegraphics[width=0.51\textwidth]{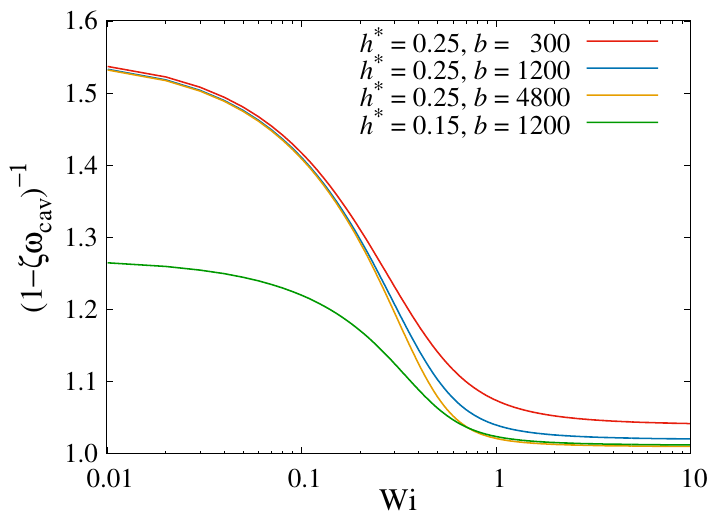}%
   \put(-199,35){\footnotesize (a)}
  \put(37,35){\footnotesize (b)}
  \includegraphics[width=0.49\textwidth]{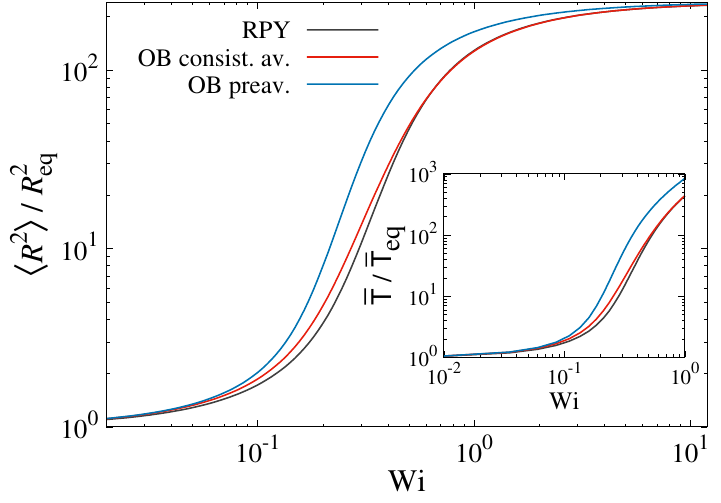}%
  \caption{(a) Variation of the prefactor in Eq.~\eqref{eqn:effWi} with Wi for different values of $b$ and $h^*$. (b) Increase of the polymer mean square extension as a function of Wi for the RPY tensor, and the preaveraged and consistently averaged versions of the OB tensor. Here $b=1200$ and $h^*=0.25$. The inset shows the mean polymer stress vs Wi for the same parameters and hydrodynamic-interaction tensors.}
  \label{fig:varpi}
\end{figure}
Now, if the prefactor $(1-\zeta\varpicav)^{-1}$ in Eq.~\eqref{eqn:effWi} were independent of Wi, then $\widetilde{\mathrm{Wi}}$ would be an effective Weissenberg number, and the effect of HI would be to rescale Wi as in the preaveraging approximation. This is not the case, though, as shown by Fig.~\ref{fig:varpi}, which plots the prefactor as a function of Wi for different values of $b$ and $h^*$. We see that the prefactor decreases strongly around the coil-stretch transition, which occurs at $\Wi = 1/2$ in the absence of HI \citep{ppv23}. It asymptotes to the constant value given by the preaveraging approximation, $(1-\zeta\varpi_{\mathrm{eq}})^{-1}$, as $\Wi\to 0$, and to a value close to unity as $\Wi\to\infty$. The latter asymptotic value approaches unity as $b$ is increased.

To better understand the effect of HI on the coil stretch transition (within the consistently averaged approximation), we can define an effective Weissenberg number $\mathrm{Wi}^\star$ using the value of the prefactor $(1-\zeta\varpicav)^{-1}$ at $\Wi = 1/2$. This is equivalent to requiring the coil-stretch transition to occur at the same value of $\mathrm{Wi}^\star$ (namely 1/2) for all values of the HI parameter $h^\star$. The variation of the prefactor near $\Wi = 1/2$, would then imply that dumbells with HI would stretch more for  $\mathrm{Wi}^\star < 1/2$ and stretch less for $\mathrm{Wi}^\star > 1/2$. Consequently, the effects of HI result in a shallower coil-stretch transition.

\begin{figure}
  \centering
  \includegraphics[width=0.5\textwidth]{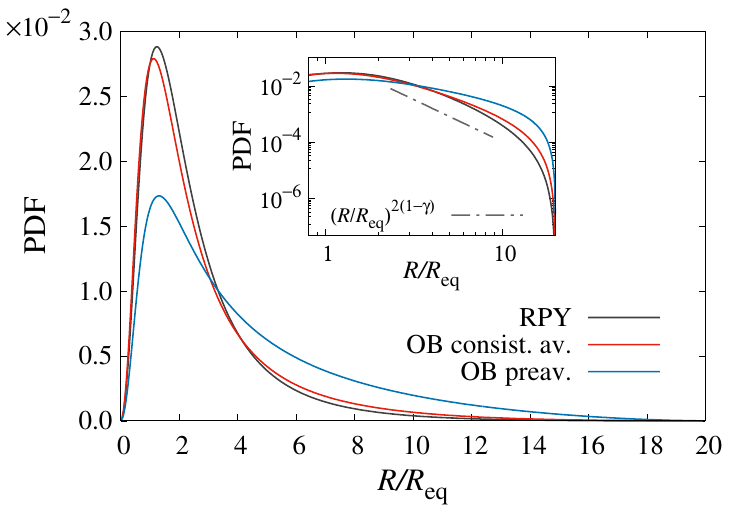}%
     \put(-190,39){\footnotesize (a)}
  \put(43,39){\footnotesize (b)}
      \includegraphics[width=0.5\textwidth]{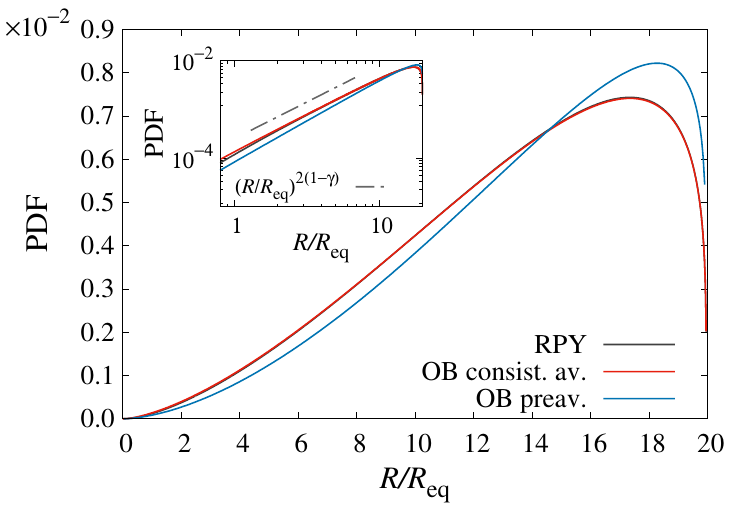}%
  \caption{(a) PDF of the rescaled polymer extension for the RPY tensor and the preaveraged and consistently averaged versions of the OB tensor. The parameters are $b=1200$, $h^*=0.25$, and $\Wi=0.3$. The inset shows the same PDFs in double logarithmic scale; the dashed line indicates the slope of the PDF for same $b$ and $\Wi$, but $h^*=0$, as given by Eq.~\eqref{eq:pdf_hstar_0}. (b) The same as panel (a) with $\Wi=3$.}
  \label{fig:pdf_varpi}
\end{figure}

We now compare the two averaging approximations with our exact results. Qualitatively, we have already seen that the consistently averaged approximation is a major improvement over the preaveraging approximation; only in the former do the effects of HI depend                                                                                                                        on Wi and $b$, decreasing when either parameter is increased.
The quantitative improvement is substantial as well, as shown in Fig.~\ref{fig:varpi}(b) which compares the mean-square extension; further comparisons of the stationary PDF of $R$ and the mean polymer stress are presented in Fig.~\ref{fig:pdf_varpi}. Here, the results for the averaged OB tensor are compared with the exact results for the RPY tensor. Such a comparison is permissible because the results of the averaged OB tensor will be the same as the averaged RPY tensor. Indeed, when the averaged OB tensor is used, the PDF of $R$ is defined on $0\leq R\leq \Rmax$, which means that the range of $R$ becomes the same for the OB and RPY tensors; in this case the two tensors have been shown, in Fig.~\ref{fig:Omega}(b), to yield near-identical results.

Since the consistently averaged approximation matches the exact results very well, we can accept the physical insight it yields: HI increases the extension of stiff dumbbells more than highly-elastic dumbbells and thereby results in a shallower coil-stretch transition.

%\begin{equation}
 % \mathfrak{G}_1 = 1 - \frac{27aR(R^2+2a^2)}{2(3R^2+4a^2)^2}, \qquad \mathfrak{G}_2 = 1 - \frac{27aR(3R^4+14a^2R^2+24a^4)}{4(3R^2+4a^2)^3}.
%\end{equation}

\section{HI effects in homogeneous and isotropic turbulence}\label{sect:dns}

We now check whether our key analytical prediction---that HI enhances the stretching of stiff dumbbells---holds in homogeneous and isotropic turbulence. We perform Brownian dynamics simulations, solving Eq.~\eqref{eq:SDE} for $\bm R(t)$ along Lagrangian trajectories in a direct numerical simulation (DNS) of a turbulent flow in a tri-periodic domain. By setting the Lyapunov exponent $\lambda$ of the Batchelor-Kraichnan flow [see Eq.~(9)] to match that of the turbulent flow, we directly compare the stretching of dumbbells in the two
flows. This allows us to quantitatively evaluate the efficacy of the Batchelor-Kraichnan flow as a model
for the turbulent stretching of dumbbells with hydrodynamics interactions.

Since we are considering the dilute limit and ignoring elastic feedback onto the flow, we can use pre-stored data of the velocity gradient $\bm\kappa(t)$ along tracer-trajectories to solve Eq.~\eqref{eq:SDE}. Specifically, we use a database of $9\times 10^5$ Lagrangian trajectories from a DNS of homogeneous isotropic incompressible turbulence (at Taylor-microscale Reynolds number $\mathit{Re}_\lambda\approx 111$), generated at ICTS, Bengaluru \citep{jr17}. The carrier flow was computed by discretizing the incompressible Navier--Stokes equations using a standard fully-dealiased pseudo-spectral method with $512^3$ grid points. Time integration was performed using a second-order Adams--Bashforth scheme. The time-trajectories of tracers were computed using a second-order Runge--Kutta method (along with trilinear interpolation to obtain the velocity and velocity gradient at the tracer's position). We calculate the Lyapunov exponent $\lambda$ (used to define Wi) from the Lagrangian trajectories 
% ($\lambda=3.66$ in simulation units) 
via the continuous $QR$ method (based on an Adams--Bashforth projected integrator along with the composite trapezoidal rule \citep{drv97}).  
% Note that we have used this database previously to study the turbulent stretching of bead-spring chains without HI \citep{ppv23}. 

The Euler-Maruyama method is used to integrate Eq.~\eqref{eq:SDE}. The velocity gradient from the Lagrangian database is available at intervals of $ \Delta t = (1.5 \times 10^{-2}) \lambda^{-1}$. We subdivide this interval into $N_\mathrm{sub}$ substeps to ensure accurate time-integration
% , especially for stiff dumbbells 
($N_\mathrm{sub} = 10$ suffices for all cases 
% the stiffest case of $Wi = 0.3$ 
considered here). 
Numerical divergence owing to the extension exceeding $\Rmax$ is prevented via the
rejection algorithm, 
proposed by \cite{o96}, which rejects those
time-step updates that yield extensions greater than $\Rmax\big(1-\sqrt{\delta t/10}\big)^{1/2}$. We have ensured that the integration time-step $\delta t =\Delta t/N_\mathrm{sub}$ is sufficiently small so that only a negligible fraction of updates are rejected over a simulation. Indeed, the constant vorticity-induced rotation and the turbulent fluctuations of the velocity gradient imply that a dumbbell will not experience continuous strong straining. Therefore, the sophisticated time-integration methods used in extensional flows are not necessary here.

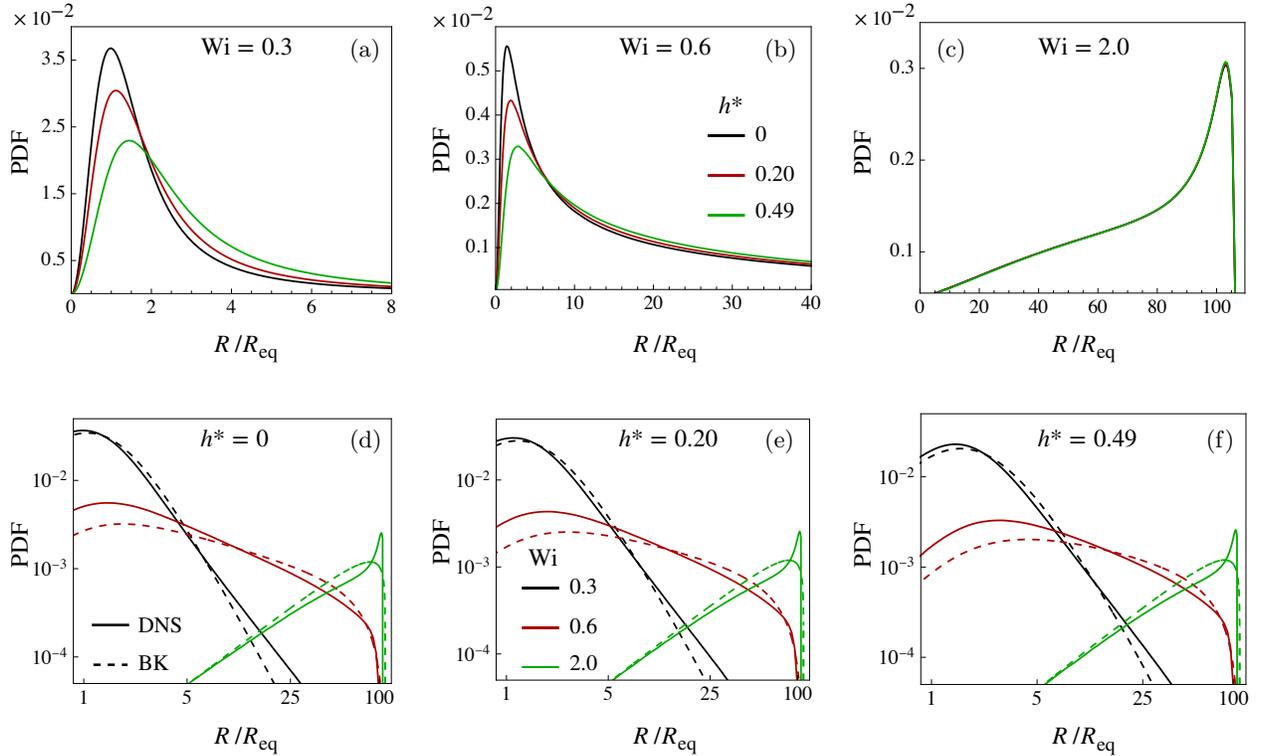
\begin{figure}
  \centering
  \includegraphics[width=\textwidth]{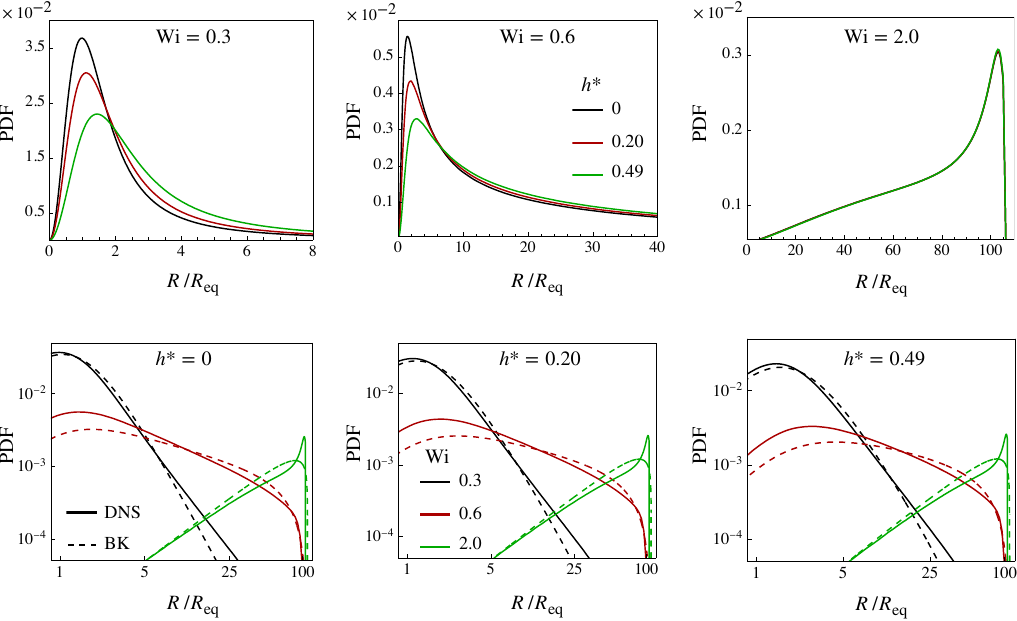}%
  \put(-340,265){\footnotesize (a)}
  \put(-183,265){\footnotesize (b)}
  \put(-118,265){\footnotesize (c)}
    \put(-340,117){\footnotesize(d)}
  \put(-183,117){\footnotesize (e)}
  \put(-16,117){\footnotesize (f)}
  \caption{(a-c) Stationary PDF of the rescaled polymer extension in homogeneous isotropic turbulence (DNS), for different values of $h^*$ (see the legend); each panel corresponds to a different value of $\Wi$. (d-f) Comparison of the PDF of the rescaled polymer extension in homogeneous isotropic turbulence (DNS) and in the Batchelor-Kraichnan (BK) flow, for different values of Wi (see the legend); each panel corresponds to a different value of $h^*$. HI are implemented using the RPY tensor, and we have set $b = 36000$ to better reveal the power-law range of the PDF of the extension. }
  \label{fig:pdf_dns}
\end{figure}

Let us first examine the effect of HI on the stationary PDF of the extension in the turbulent flow. Figs.~\ref{fig:pdf_dns}(a-c) present the PDFs for three values of $h^*$, for Wi = 0.3, 0.6, and 2.0, respectively. We see that, just as predicted by the Batchelor-Kraichnan model, HI increases the stretching of small-Wi dumbbells [compare Figs.~\ref{fig:pdf_dns}(a) and \ref{fig:pdf_RPY}(a)]. Of course, the effect of HI becomes negligible at large Wi, when the majority of dumbbells are highly extended [Fig.~\ref{fig:pdf_dns}(c)].

Next, let us compare the analytical predictions from the Batchelor-Kraichnan (BK) model with the simulation results for the turbulent flow (DNS). The corresponding PDFs of extension are compared, for three different values of $h^*$, in Fig.~\ref{fig:pdf_dns}(d-f). We see good qualitative agreement between the results of the two flows, with the exception of the local peak near $\Rmax$ in the high-Wi PDF (Wi = 2.0), that is exhibited only for the turbulent flow. This local peak was shown, by \cite{mv11}, to be a consequence of the velocity gradient having a finite correlation time, as is the case in turbulent flows. (Typically, this correlation-time is of the order of the Kolmogorov time-scale \citep{y01}, which in turn is about one-tenth $\lambda^{-1}$ \citep{bbbcmt06}.) In a correlated flow, a dumbbell can experience sustained high-straining that maintains their extension near $\Rmax$, for a while, before inevitable fluctuations in the strain-rate allow the dumbbell to relax. Such persistent straining events are, of course, not possible in the delta-correlated Batchelor-Kraichnan flow. 

The other distinctive feature of turbulent flows is the prevalence of a flared-tail, strongly non-Gaussian, distribution of strain-rates \citep{Schumacher2012,Ishihara2007,Yeung2012,Buaria2019}. In the absence of HI, we have shown that these extreme events cause small-Wi dumbbells to stretch more in turbulence than in the Gaussian Batchelor-Kraichnan flow \citep{ppv23}. This is illustrated by the PDFs for Wi = 0.3 in Fig.~\ref{fig:pdf_dns}(d) (solid and dashed black lines); we see more highly stretched polymers in the DNS. Interestingly, higher Wi polymers stretch more in the Batchelor-Kraichnan flow, as evidenced by the PDFs for Wi = 0.6 in Fig.~\ref{fig:pdf_dns}(d) (solid and dashed red lines). As explained in \citep{ppv23}, the Batchelor-Kraichnan flow (constructed to have the same Lyapunov exponent as the turbulent flow) compensates for the lack of extreme strain-rates by a greater abundance of moderate strain-rates. For dumbbells with a sufficiently large relaxation time, the cumulative effect of moderate straining produces larger extensions than short-lived bursts of extreme straining. Does this picture hold even with HI? The answer is yes, as demonstrated by Figs.~\ref{fig:pdf_dns}(e-f), which present results for $h^*=0.2$ and 0.49 that are entirely analogous to those for $h^* = 0$ [Fig.~\ref{fig:pdf_dns}(d)]. 

So, the Batchelor-Kraichnan flow, which is known to capture the qualitative features of turbulent dumbbell stretching without HI \citep{ppv23}, serves as a useful model even when HI are introduced into the dumbbell. Indeed, our analytical predictions of HI effects---made possible by adopting the Batchelor-Kraichnan model---are borne out in the results of simulations in homogeneous and isotropic turbulence.

\section{Concluding remarks}
\label{sect:conclusions}

The simple dumbbell model captures the essential two-way interaction between the elastic microstructure and the macroscale flow in a polymer solution: fluid drag acts on the beads of the dumbbell and stretches it out, even as the beads exert a net back reaction force---of the same magnitude as the elastic spring force---and modify the flow. This elastic feedback has two consequences: (i) a large-scale modification of the flow field that, given a sufficient number of dumbbells, produces effects like turbulent drag-reduction \citep{g14,x19} and elastic turbulence \citep{steinberg21}; (ii) a microscale effect associated with the alteration of the flow field at the scale of an individual dumbbell. The latter effect, which is that of hydrodynamic interactions, can be studied separately by considering the dilute limit---as we do here---in which the large-scale flow is unaffected by the dumbbells. In most applications, though, both effects are present. 

Now, nearly all simulations of complex viscoelastic flows are based on the dumbell model, either indirectly via continuum constitutive equations like Oldroyd-B and FENE-P, or directly as in Eulerian-Lagrangian simulations that evolve a large number of dumbbells in a flow field computed from the Navier-Stokes equations \citep{Keunings04}. The second multiscale approach is promising because of its ability to easily represent molecular-scale phenomena, such as finite extensibility \citep{serafini23} and mechanical scission \citep{s20,vwrp21}. However, these simulations typically do not account for HI. Techniques based on the CONFESSIT approach \citep{connffessit} use the dumbbells within a grid cell to construct the polymer stress tensor \citep{Keunings04,wg13,vwrp21}; the elastic force that is then obtained by taking the divergence of the stress tensor acts only on the macroscale flow. Thus, unless HI are explicitly incorporated into the evolution equation for the dumbbells, their effect is lost. And HI are usually neglected because of the computational cost of evolving the very large number of dumbbells needed to accurately compute the stress tensor. The alternative approach to multiscale simulations is to solve for the disturbance flow created by the beads of the dumbbells exerting (regularized) point-forces onto the fluid \citep{sbgc22,rll22}. In principle, this approach should naturally capture the effects of HI; however, this would require an extremely fine spatial resolution, on the order of the polymer length scale, which is not practical in most circumstances. So, the question arises: What is the effect of the neglected HI on the accuracy of multiscale dumbbell simulations?

Our study answers this question for homogeneous isotropic turbulence. Our analytical results---confirmed by simulations in a turbulent flow---show that HI only mildly increases the extension of small-Wi dumbbells. The effect on the averaged extension, and thereby on the mean polymer stress, is quite small. Moreover, the effects of HI vanish at high-Wi, which is the case of interest in most applications. Hence, our results suggest that one may neglect HI in multiscale dumbbell simulations, especially given the other physical assumptions associated with representing polymers by dumbbells.

Will the elastic modification of the large-scale flow alter the impact of HI? From the perspective of an individual dumbbell, the statistics of the velocity gradients it experiences will be altered (e.g., in high-Re homogeneous isotropic turbulence, elastic forces dampen the extreme strain-rates \citep{prasad10,rll22}). However, the change in the average strain-rate can be accommodated by rescaling Wi, while alterations to the distribution of strain-rates has been shown here to have no impact on the qualitative effects of HI---our results for the Gaussian Batchelor-Kraichnan flow exhibit the same HI effects as those for the non-Gaussian turbulent flow. Therefore, we believe that our conclusions regarding the influence of HI are relevant even for viscoelastic flows with elastic feedback effects.

Apart from its practical implications, our analytical study has helped us gain insight into the stretching of dumbbells with HI in a fluctuating flow. The consistently-averaged approximation shows that the effects of HI cannot be captured by a simple rescaling of Wi. 
% While HI causes small-Wi dumbells to stretch more, it has little effect on large-Wi dumbbells. As a consequence,
Rather, the Wi-dependence of the effects of HI results in a less pronounced coil-stretch transition in fluctuating chaotic flows.

The natural next step in the investigation of the effects of HI on polymer stretching in turbulent flows is to study the stretching of bead-spring chains. The presence of a larger number of beads, as well as the configurational degrees of freedom of a chain, could result in stronger effects of HI. We hope to make progress in this direction in the near future. 

\acknowledgments

We thank Samriddhi Sankar Ray (ICTS, Bengaluru) for sharing his database of Lagrangian trajectories in homogeneous and isotropic turbulence. This work was supported by the Indo–French Centre for the Promotion of Advanced Scientific Research (IFCPAR/CEFIPRA, project no. 6704-1). Both authors acknowledge their Associateships with the International Centre for Theoretical Sciences (ICTS), Tata Institute of Fundamental Research, Bengaluru,
India. D.V. acknowledges the support of Agence Nationale de la Recherche through Project
No.~ANR-21-CE30-0040-01.
Simulations were performed on the IIT Bombay workstation \textit{Aragorn} (procured through the IIT-B grant RD/0519-IRCCSH0-021).

\bibliography{polymers}

\begin{thebibliography}{66}
\bibitem [{not()}]{note-equilibrium}%
  \BibitemOpen
  \href@noop {} {}\bibinfo {note} {If $\bm\kappa=0$, th PDF of $\bm R$ is independent of the form of $\boldsymbol{\mathsf{\Omega}}$ and coincides with the equilibrium PDF for a free-draining dumbell. Indeed, $\partial D_{ij}/\partial {R_j}=0$, $D_{ij}^{-1}V_j=-F_i/2k_BT$, and $\bm F$ is the gradient of a scalar potential. Therefore, Eq.~\eqref{eq:FPE-laminar} admits the stationary zero-probability-current solution $\psieq(\bm R) \propto \exp[-\Phi(R)/k_bT]$ with $\Phi(R)=(\int_0^{R} \vert\bm F\vert\, dR')/k_bT$, where we have used $\bm F=\vert\bm F\vert\bm R/R$ (see \cite{gardiner}, Sect.~5.3.3).}\BibitemShut {Stop}%
\end{thebibliography}

\end{document}